# Plastic strain-induced olivine-ringwoodite phase transformation at room temperature: main rules and the mechanism of the deep-focus earthquake

Feng Lin[1]*, Valery I. Levitas[1, 2, 3]*, Sorb Yesudhas[1], Jesse Smith[4]

[1]Department of Aerospace Engineering, Iowa State University, Ames, Iowa 50011, USA

[2]Department of Mechanical Engineering, Iowa State University, Ames, Iowa 50011, USA

[3]Ames National Laboratory, US Department of Energy, Iowa State University, Ames, Iowa 50011, USA

[4] HPCAT, X-ray Science Division, Argonne National Laboratory, Argonne, Illinois 60439, USA

*Corresponding authors. Email: flin1@iastate.edu and vlevitas@iastate.edu

Deep-focus earthquakes that occur at 350–660 km are theorized to be caused by strain-induced olivine→spinel phase transformation (PT). We introduce and apply dynamic rotational diamond anvil cell with rough diamond anvils to deform San Carlos olivine. While olivine was never transformed to spinel at any pressure at room temperature, we obtained olivine-ringwoodite PT under severe plastic shear at 15-28 GPa within seconds. This is conceptual proof of the difference between pressure- and plastic strain-induced PTs and that plastic straining can accelerate this PT from million years to timescales relevant for the earthquake. The PT pressure linearly reduces with increasing plastic strain, corresponding increasing dislocation density and decreasing crystallite size. The main rules of the coupled severe plastic flow, PT, and microstructure evolution are found.

**One-Sentence Summary:** Olivine-ringwoodite phase transformation was obtained under severe plastic shear at 15-28 GPa and room temperature within seconds, which conceptually proves it as a possible mechanism for deep-focus earthquakes.

Deep-focus earthquakes occur at 350–600 km at pressures of 12-23 GPa and temperatures of 900-2000 K and cannot be explained by brittle fracture, in contrast to shallow earthquakes [1]. The main hypothesis is that the deep-focus earthquakes are caused by shear localization due to PT from the subducted metastable olivine (forsterite) to denser β-spinel (wadsleyite) or γ-spinel (ringwoodite) [2-12]. However, there are many existing enigmas, which particularly were theoretically addressed in [13], but without even conceptual experimental confirmation. The main enigma is: How does metastable olivine, which does not completely transform to spinel deeply in the region of stability of spinel for over a million years, suddenly transform within shear bands during seconds to produce a seismic strain rate of $10\text{-}10^3/s$? One of the main points in [13] is that the olivine-spinel PT is treated as plastic strain-induced (instead of pressure/stress-induced), which leads to completely different thermodynamic and kinetic treatments [14-17]. For strain-induced PTs, the PT rate is proportional to the strain rate (Eq. (1)), i.e., high strain rates lead to high PT rates. However, there was not any experimental proof of the existence of strain-induced mechanisms for olivine-spinel PT. In experiments involving plastic deformation [18, 19], it is relatively small (~0.3), and a temperature above 1100 K is required for PT to occur within hours; these PTs are treated as pressure/temperature-induced ones. Moreover, most experimental confirmations of PT-based mechanisms were obtained for structural analogs of olivine, $Mg_2GeO_4$ [4, 6, 10-12], or $Fe_2SiO_4$ [3], in which PT occurs at much lower pressure and temperature. In this case, a brittle fracture may occur first, and friction at the shear crack surfaces may cause PT [11] rather than vice versa. Here, we conceptually prove that indeed severe plastic shear, even at room temperature, can produce the olivine-spinel PT within seconds in the Mg-rich San Carlos olivine (($Mg_{0.91}Fe_{0.09})_2SiO_4$) in the pressure range 15-28 GPa, estimated plastic strain 2.3-9, at strain rate $3.5\times10^{-3}$-$3.3\times10^{-2}$/s, while this PT was not reported under hydrostatic and nonhydrostatic compressions at any pressure. For this purpose, we introduce dynamic rotational diamond anvil cell (dRDAC), which allowed us to apply a controllable rotation rate, and rough diamond anvils (rough-DA), which increase the friction shear stress at sample-anvil boundary up to a maximum value equal to the yield strength in shear, minimizing relative sliding and intensifying shear flow and localization. Radial distributions of pressure, the volume fraction of ringwoodite, dislocation density, and crystallite size, all averaged over the sample thickness, were determined for different compression-shear loadings, thus generating plentiful data from a single experiment. Olivine transforms directly to ringwoodite without wadsleyite. The first rules were found that the minimum PT pressure linearly reduces with increasing plastic strain and corresponding increasing dislocation density and decreasing crystallite size. This explains the necessity of severe plastic strain and nanograined olivine structure for strain-induced olivine-ringwoodite PT, which is easily achieved in geodynamic activity but not in any previous experiment. The maximum volume fraction of ringwoodite does not exceed 0.4, which, combined with theoretical/computational results [14, 17, 20], implies the existence of shear-transformation bands in a sample, the required condition for the deep-focus earthquake [13]. The crystallite sizes of the olivine and ringwoodite are steady and independent of plastic strain, strain path, pressure, and the volume fraction of ringwoodite. The same is true for the dislocation density in the olivine but below some plastic strain magnitude. The crystallite size of ringwoodite is 5-10 nm, consistent with the nucleation at the tip of the dislocation pileups mechanism without further growth [14-17].

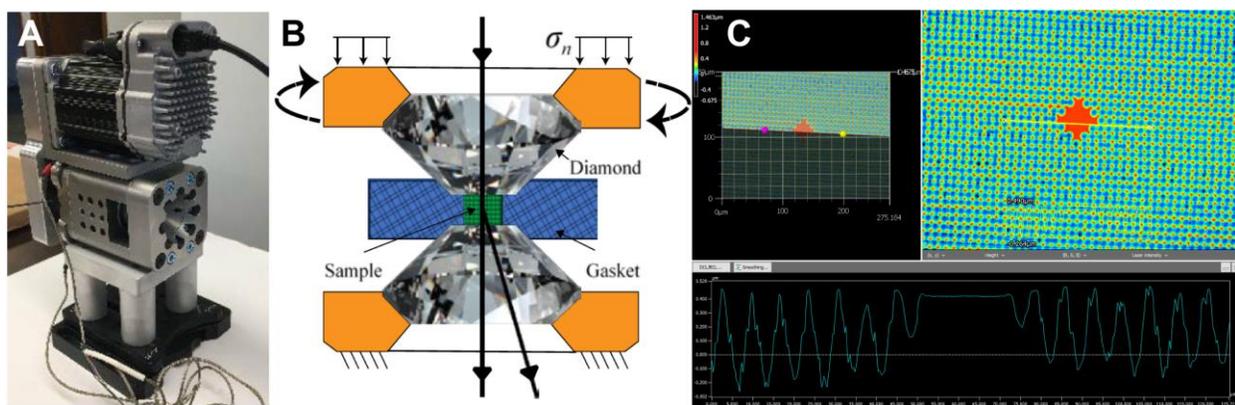

**Fig. 1. Dynamic rotational diamond anvil cell (dRDAC) with rough diamond anvils.** **(A)** Design of dRDAC (DAC Tools LCC, Dr. Stanislav Sinogeikin); **(B)** Schematic of dRDAC; **(C)** Periodic asperity profiles at the surface of diamond anvil produced by laser ablation (PALM-Scientific LLC, Dr. Sergey Antipov).

The basic difference between PTs under hydrostatic and non-hydrostatic compression below the yield (pressure- and stress-induced PTs) and PTs during plastic flow (strain-induced PTs) was formulated in [14] and elaborated in [15-17, 21-23]. The pressure- and stress-induced PTs are initiated by nucleation at the pre-existing defects, like dislocations or grain boundaries. Conversely, the *plastic strain-induced PTs start at new defects, like dislocation pileups, continuously generated during the plastic flow*. This is the only *mechanism* that can explain the reduction of the PT pressure by one to two orders of magnitude observed in experiments for various materials [22-25]. The concentration of all stress components at the tip of a dislocation pileup is proportional to the number of dislocations $N$ in a pileup; since $N$ can be large, from 10 to 100, local stresses are greatly multiplied (much larger than for stress-induced PT), which drastically reduces the external pressure required for PT. Also, the kinetics of strain-induced PTs has the following structure [14]:

$$\frac{dc}{dq} = f(p,q,c) \rightarrow \frac{dc}{dt} = f(p,q,c)\frac{dq}{dt} \qquad (1)$$

where $c$ is the volume fraction of the high-pressure phase, $q$ is the accumulated plastic strain, and $p$ is the pressure. Time $t$ is not a governing parameter in Eq. (1), but plastic strain plays a role of a time-like parameter. Consequently, the PT rate is proportional to the plastic strain rate, which was utilized in [13] to explain the drastic increase of the olivine-spinel PT rate in a shear band during the earthquake. We obtained here this PT within seconds utilizing this property.

A new dRDAC is utilized (Fig. 1), which produces the rotation of an anvil with a controllable rate in a broad range. Based on previous experience with other materials (Zr [21, 22] and Si [23]) and theoretical predictions [14-17, 22, 23], to maximally reduce PT pressure, the maximized plastic strain with promoted dislocation density is needed. Due to relative sliding between the sample and the anvil, the rotation of the sample may be 5 times smaller than that of the anvil [24]. That is why the dRDAC and rough-DAs with the periodic asperity height of 300 nm (instead of 10 nm for standard anvils), produced by laser ablation, are implemented. Asperities penetrate the sample and gasket surface, and sliding occurs within deformed materials. Contact friction reaches its maximum possible magnitude equal to the pressure-dependent yield strength in shear, which minimizes sliding and promotes plastic flow and PT.

San Carlos Olivine ($Mg_{0.91}Fe_{0.09}SiO_4$) powder (see Table S1 for detailed element analysis) was compacted in an iron gasket with a chamber diameter of 200 µm and an indented thickness of ~150 µm. The sample is compressed and sheared along the path defined in Table S2. At each torsion step, synchrotron diffraction images were taken along the sample radius with a wavelength of 0.3445 A and 10 µm step size. Pressure was calibrated using $3^{rd}$ Birch-Murnaghan equation of state of San Carlos olivine from [26]. Fig. 2 compares examples of X-ray diffraction patterns without and with PT at rotation angles $\phi=180°$ and $270°$. No PT was observed at $\phi=180°$. After imposing another 90° of rotation to $\phi=270°$, pressure increases significantly with a radial pressure gradient (Fig. 3A). Ringwoodite phase was first time observed at $\phi=270°$ with the maximum c=0.4 at sample center and decreasing down to zero towards the edge (Fig. 3B). Note that wadsleyite (which appear between olivine and ringwoodite under hydrostatic compression) was not observed in the experiment; thus, PT sequence for strain-induced and pressure-induced olivine-spinel PTs is different. With further rotation to 300° and 330°, the pressure distribution is stabilized and does not change essentially. The local pressure disturbance is due to volume reduction and stress release during PT. Although it is not feasible to experimentally track deformation path of each material particle within the sample, the accumulated plastic strain $q$ at the given sample thickness $h$ and radius $r$ with respect to the culet center can be approximately estimated as a sum of compression and shear strains using [27]:

$$q = \ln\left(\frac{h}{h_0}\right) + \frac{\phi r}{\sqrt{3}h} \quad (2)$$

where $h_0$ is the sample's initial thickness, $\phi$ is rotation angle in radian. Since shear strain is imposed at different thicknesses $h$, this equation is used incrementally between each step along the deformation path and combining all increments. Comparing the evolution of the volume fraction of ringwoodite from $\phi$=270º to 330º rotation at the same radii, where pressures are very close, but shears increase with torsion, we see that the volume fraction $c$ increases with increasing $q$, which is consistent with the plastic strain controlled kinetic Eq. (1). After rotation of the anvil stops, the volume fraction of ringwoodite stops evolving as well, i.e., time is not a governing parameter in the PT kinetic, like in Eq. (1). Obtained results provide conceptual proof of the validity of strain-controlled kinetics for olivine-ringwoodite PT. Also, with a rotation rate of 0.1 RPM, 30º increment takes 50 s, which is a characteristic time for PT in the current experiment. Since time is not a parameter in kinetic Eq. (1), increasing the rotation rate by one-two orders of magnitude would reduce the characteristic PT time proportionally to 5 and 0.5 s. Thus, utilizing strain-induced PT versus pressure-induced PT reduces the PT time from million years down to seconds, which is required for the PT-based mechanism of the deep-focus earthquake [13].

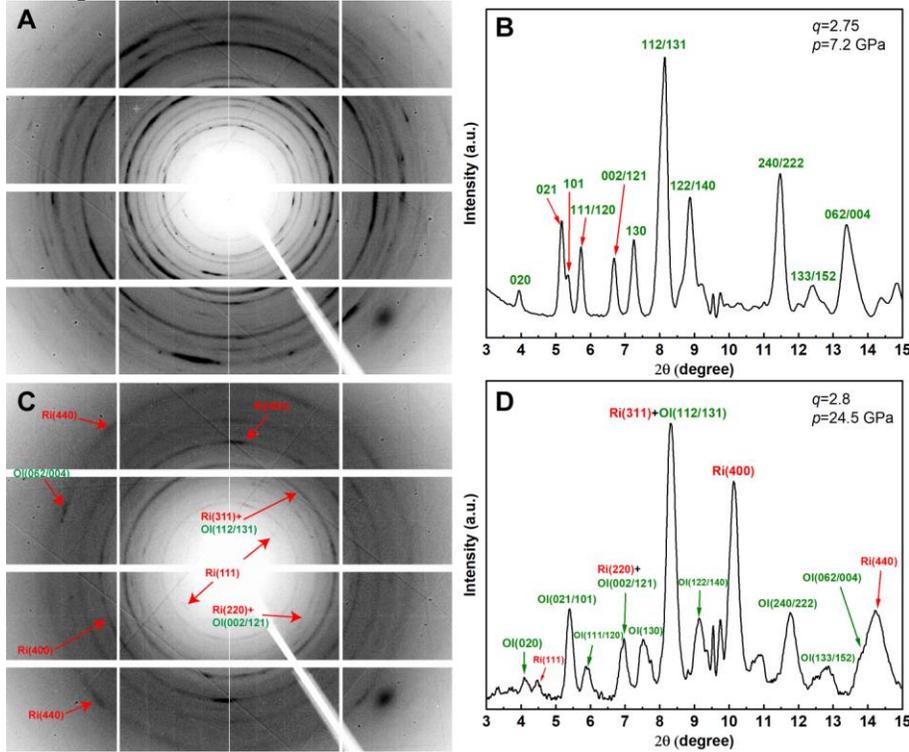

**Fig. 2. X-ray diffraction patterns of deformed olivine sample.** (**A**) 2D patterns and (**B**) integrated peak profiles after olivine reaching $p$=7.2 GPa at the sample center and $180^0$ of anvil rotation. (**C**) 2D patterns and (**D**) integrated peak profiles of olivine and ringwoodite after reaching $p$=24.5 GPa at the sample center and $270^0$ of anvil rotation.

The same maximum c≈0.4 for $\phi$=300º and 330º near the sample center means that the volume fraction reached the steady value for the given pressure. According to the theory [14], one of the necessary conditions for the steady solution to Eq. (1) is that plastic strain simultaneously promotes direct and reverse PTs in different regions of the representative volume. However, this should not be the case here. Indeed, the crystallite size of the ringwoodite is ~5-10 nm (Fig. 3D), consistent with the grain size of nucleated ringwoodite of 10-20 nm obtained during plastic compression of olivine at high temperature [19]; it grows with time to 100 nm leading to the increase in the yield strength by a factor of 4. This confirms that 10 nm grains are deeply in the region of the inverse Hall-Petch rule for the yield strength, and grain boundary sliding is expected. Consequently, no significant dislocation activity with large dislocation pileups in the ringwoodite to promote reverse PT is expected. In this case, kinetic Eq. (1) cannot have a steady solution [14], and complete transformation to ringwoodite is expected with increasing plastic strain, as it occurs in other materials [21, 22, 28-30]. Even if the reverse PT would occur, a steady volume fraction is possible in the case of a very strong effect of plastic strain on it, namely when the reverse PT can start at pressures higher than the pressure for the direct strain-induced PT [14], but this is not documented for any material and not the case here.

The only currently reasonable explanation of the steady volume fraction is an assumption that plastic strain and strain-induced PT are localized in the shear bands and do not essentially spread outside of them. This is qualitatively expected since the yield strength of the nanograined ringwoodite is 4-6 times lower than that of the olivine [19]. Finite element solution for compression and torsion in RDAC of the model material with the yield strength of the high-pressure phase 3 times lower than that for the low-pressure phase within a gasket [18] shows a system of the intersecting shear-PT bands (Fig. 3E-G), one of which is localized at the contact surface and other are inclined to the symmetry axis. After some torsion, the PT completes in each band, and further evolution of volume fraction is limited; thus, the volume fraction of the high-pressure phase averaged over the thickness remains approximately constant. Evaluating visually the radial distribution of the volume fraction averaged over the thickness in Fig. 3E, we obtain qualitatively the same non-monotonous features as observed in the experiment in Fig. 3B. This contrasts with monotonous radial volume fraction distribution for materials with stronger high-pressure phase and no shear banding, like in Zr [21, 22] and BN [28]. While comparing curves $c(r)$ in Fig. 3B for different rotation angles, we can interpret the jumps in the volume fraction in terms of the appearance of a new shear-PT band and a small increase in $c$ within the same bands. Unfortunately, we could not check the existence of the shear-PT bands on the retrieved sample because it pulverized during the unloading due to severe plastic strains and large internal stresses. We also could not perform more quantitative finite element simulations for the olivine-spinel system because PT kinetics and pressure-, plastic strain-, grain size-, and dislocation density-dependent yield strength of phases during severe plastic flow are unknown and should be determined via coupled experimental-computational procedure [31, 32], which will be done in future. However, this was never done for a sample with shear bands, which significantly complicates the problem.

To reveal the main rules of the microstructure evolution and its relation to plastic strain and strain-induced PT, we in-situ track the evolution of the radial distribution of the crystallite size d and dislocation density $\rho$ averaged over the sample thickness measured from synchrotron XRD (Fig. 3C and D). Crystallite size is extracted from Rietveld refinement using Material Analysis Using Diffraction (MAUD) software [33], and dislocation density is estimated using the William-Smallman method [34] using the easiest slip system (010)[001] [35] (see supplementary for details). At $\phi$=180º, the crystallite size is relatively large, and dislocation density is relatively low, around 40-60 nm and $2.5 \times 10^{14} m^{-2}$, respectively, which leads to an insufficient stress concentrator to trigger PT at a low pressure of 6-8 GPa. The crystallite size of olivine significantly decreases to ~25 nm, and dislocation density increases to ~$5.5 \times 10^{14} m^{-2}$ after compression to ~24 GPa at the sample center and ~14 GPa at the edge, followed by 270º rotation and some PT. The next two rotation increments do not essentially change the distribution of the crystallite size in olivine along the radius (Fig. 3D). Considering statistical scatter and including all data for $\phi$=270º-330º and all radii, we may conclude that the crystallite size is practically the same. However, the pressure, plastic strain, and volume fraction of the ringwoodite are very different along the radius and during the torsion increments. The plastic strain path is also very different: uniaxial compression at the center with increasing shear along the radius. Thus, the obtained result implies a new, very informative rule of coupled severe plastic flow, strain-induced PT, and microstructure evolution: *after severe plastic straining under high pressure, an averaged crystallite size in the olivine is getting steady and independent of plastic strain, and its path, pressure, and volume fraction of the ringwoodite.*

At the same time, after $\phi$=180º at 6-8 GPa before PT, the scatter in the crystallite size is much larger, and no statistically strict conclusions can be made. For the same loading, dislocation density distribution has a small scatter and is practically constant along the radius (Fig. 3C). While pressure variation is small, we still can conclude that *after severe plastic straining before PT, dislocation density is getting steady and independent of plastic strain and its path*. Of course, this should be confirmed for plastic straining at higher pressures before PT. During PT at $\phi$=270º and 300º, *dislocation density can also be considered practically constant and independent of radius, i.e., independent of plastic strain and its path, pressure, and volume fraction of the ringwoodite*. However, at $\phi$=330º, scatter significantly increases, and no conclusions can be made. This may be caused by heterogenous processes due to shear strain and PT in the bands, which lacks the understanding of our knowledge and is the main challenge for future studies. The crystallite size of ringwoodite is small, around 5-10 nm (Fig. 3D), and practically independent of radius. Thus, *an averaged crystallite size in the ringwoodite during the strain-induced PT is steady and independent of plastic strain and its path, pressure, and volume fraction of the ringwoodite.* The small crystallite size was predicted and expected for nucleation at the tip of dislocation pileup [14] because all stresses reduce fast away from the tip, and growth is arrested. Next ringwoodite crystal appears during the next small plastic strain increment leading to new dislocation pileups.

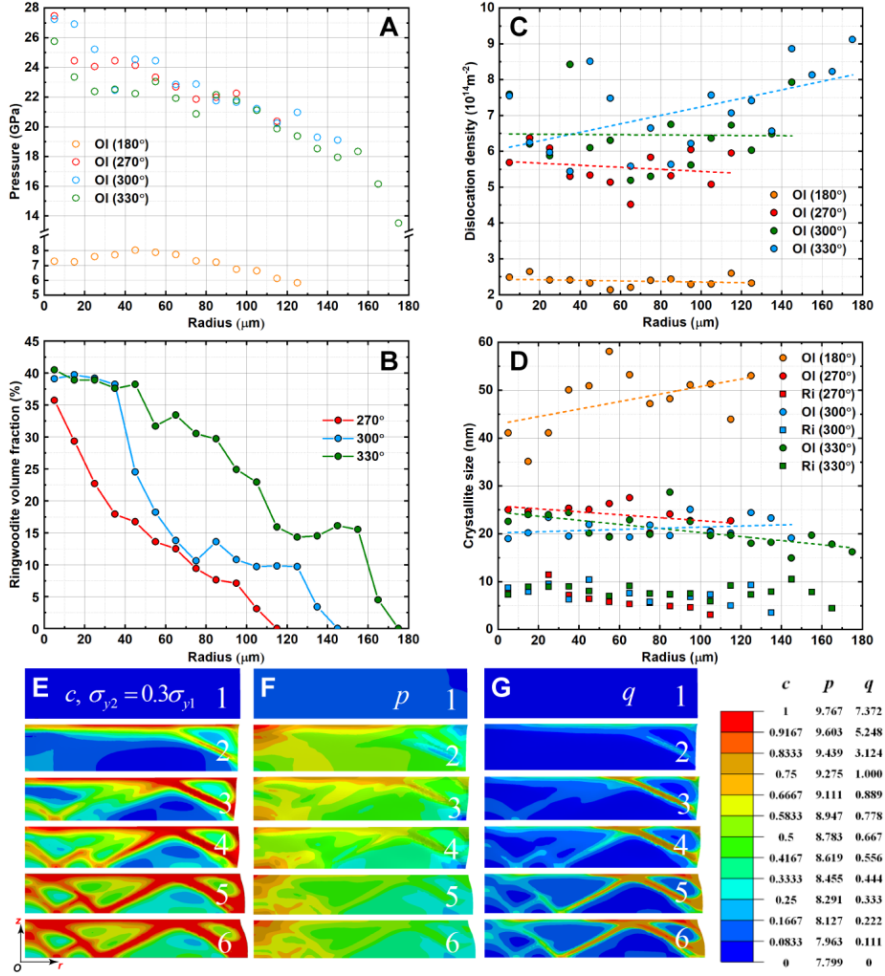

**Fig. 3. Radial distributions of various parameters in the samples after different compression-torsion loadings in dRDAC.** (A) Pressure; (B) volume fraction of the ringwoodite; (C) dislocation density; and (D) crystallite size. Finite element simulation results [18] showing distributions of (E) volume fraction of the high-pressure phase $c$; (F) pressure $p$; and (G) accumulated plastic strain $q$ in the sample with 3 times lower yield strength of the high-pressure phase than that of the low-pressure phase, within a gasket (not shown) under a constant compressive force with increasing rotation angle $\varphi$ of (1) 0, (2) 0.1, (3) 0.3, (4) 0.5, (5) 0.8, and (6) 1.0 radians.

The minimum PT pressure is plotted as a function of crystallite size and dislocation density of olivine and estimated plastic strain in Fig. 4 using only the data around the edge where PT just starts ($c<0.1$). The minimum PT pressure decreases with increasing dislocation density, plastic strain, and decreasing crystallite size, demonstrating a linear relationship within the experimental range with the following approximation:

$p_{min}(GPa) = -2.04\rho_{ol} + 32.52;$   $\rho_{ol}$ is dislocation density of olivine in $(10^{14} m^{-2})$   (3)
$p_{min}(GPa) = 1.13 d_{ol} - 4.68;$   $d_{ol}$ is the crystallite size of olivine in $(nm)$   (4)
$p_{min}(GPa) = -1.31q + 27.01;$   $q$ is the accumulated plastic strain   (5)
$d_{ol} = -1.16q + 28.04$   (6)
$\rho_{ol} = 0.64q + 2.70$   (7)

We can hypothesize the following interpretation of the trends in Eqs. (3)-(5). All components of the stress tensor, $\sigma$, at the tip of the edge dislocation pileup, are $\sigma \sim \tau l$ [36]; here $\tau$ is the applied shear stress limited by the yield strength in shear $\tau_y$, and $l$ is the dislocation pileup's length. Since $l$ is usually taken as a fraction of the grain size $d$ (e.g., $l=0.5d$), the main implication in [14] was that the larger the grain size, the stronger the stress concentration for all stresses and, consequently, the larger the reduction in transformation pressure. This is opposite to what is found in the current experiments and for PT in Zr [21, 22] and Si [23]. To resolve this paradox, we utilize that in the phase field [15,16], molecular dynamics [37], and concurrent atomistic-continuum simulations [38, 39], almost all dislocations are concentrated at the grain boundary producing a superdislocation or step (Fig. S1) with effective length $l=Nb<<d$, where $b$ is the magnitude of the Burgers vector, and $N$ is the number of dislocations in a pileup. The number of dislocations equilibrated in a superdislocation $N \sim \tau = \tau_y$ [38, 39]. At the same time, combining the Hall-Petch contribution due to grain size $d$ and the Taylor contribution due to dislocation density $\rho$ [40], one obtains:

$$\tau = \tau_y = \tau_0 + k d^{-0.5} + m \rho^{0.5} \qquad (8)$$

where $\tau_0$, $m$, and $k$ are material parameters. That is why the stress concentration increases and the minimum pressure for the strain-induced PT decreases with decreasing grain size (which resolves the above paradox) and increasing dislocation density. Also, the higher the dislocation density, the higher the probability of generating more dislocation pileups with a larger number of dislocations, reducing the PT pressure. Lower grain size (a) reduces chances for a dislocation forest, which may reduce the probability of the appearance of dislocation pileups, and (b) increases misorientation between grains, which increases resistance for dislocations to pass through the grain boundary. Plastic strain is the tool for increasing the dislocation density, reducing crystallite size, and bringing up more dislocation pileups with large $N$. The obtained results show that with a significant increase in the plastic strain, the PT pressure can be potentially further reduced below the current minimum value of 15 GPa (which is already lower than the extrapolated phase equilibrium pressure between 15 and 16 GPa for wadsleyite-ringwoodite at 300 K [41]).

One of the seeming contradictions is that the crystallite size in olivine reaches a steady value in Fig. 3D, but we found a reduction in PT pressure with the reduction in the crystallite size in Fig. 4B. However, the steady state is claimed for the averaged values for all data in the statistical sense. At the same time, nucleation is determined by the strongest stress concentrator. For points corresponding to initiation PT, crystallite size reduces from 22 to 17 nm during rotation angle increment from 270º to 330º.

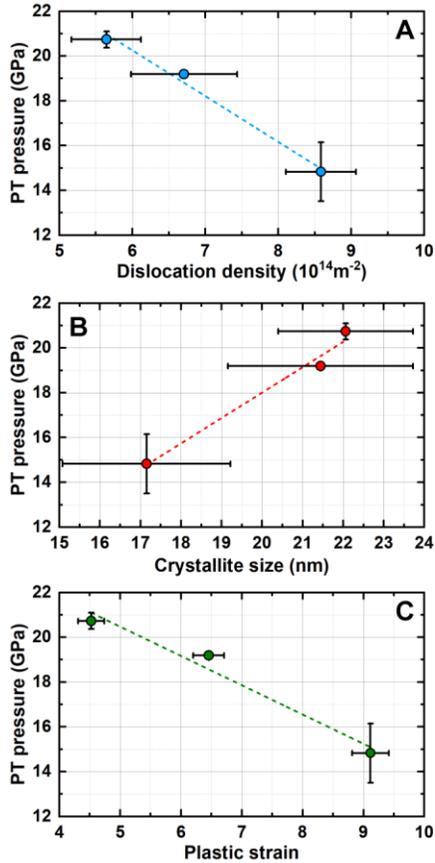

**Fig. 4.** Minimum pressure for initiation of strain-induced PT versus (**A**) dislocation density; (**B**) crystallite size; and (**C**) plastic strain.

In summary, with newly introduced dRDAC, rough-DA, and measurements of the main transformational and microstructural parameters along the sample radius, we received a conceptual proof of the plastic strain-induced olivine-ringwoodite PT even at room temperature within seconds, and the main rules for this PT and microstructure evolution. The key to success in the reduction in PT pressure was to increase dislocation density and reduce the crystallite size by severe plastic flow. The crystallite size of ringwoodite is as small as 5-10 nm, consistent with the dislocation pileup-based mechanism for strain-induced PTs. Corresponding plastic strain, $q$=2.3-8.8, can be reached at the boundary of the subducting olivine wedge, especially within shear bands, but was not obtained in previous compression experiments [42]. Similar strain-induced PT is expected at temperatures of 900-1200 K, relevant to the deep-focus earthquakes [13, 18, 19]. Indeed, in nonhydrostatic DAC experiments [43] on $Mg_2SiO_4$, the thin β-spinel film appeared at the diamond surface at 19-35 GPa and holding at 575°C for 10.5 hours. No PT was found in bulk, which indicates that this PT is induced by large plastic shears due to friction in the contact region. Even at 28-32 GPa and 650°C for 8 hours, wadsleyite was observed in bulk but not ringwoodite. Much larger accumulated shear strains provided in dRDAC should further reduce PT pressure at high temperatures and promote PT kinetics. Obtained results open windows for coupled experimental/theoretical/computational studies of plastic strain-induced olivine-ringwoodite PT and shear localization in a broad range of strain rates and temperatures with more realistic applications to the deep-focus earthquakes problems and larger-scale processes during olivine slab subduction [44, 45]. In particular, severe transformation-induced plasticity and corresponding heating in the shear-PT band predicted in [13] can be confirmed by combining experiments with simulation of sample behavior, like in Fig. 3. Our results also may lead to reconsideration of numerous known PTs in Earth as plastic strain-induced instead of pressure-induced, which may change many geological interpretations. For example, they may explain the appearance of microdiamonds directly in the cold Earth crust within shear bands [25] without subduction to the high-pressure, high-temperature mantle and uplifting. We want to stress that the main challenge for nonhydrostatic experiments—strongly nonuniform fields—can be transformed into a remarkable opportunity to generate big data with various material characterizations, like in this study. In addition, in high-pressure torsion processing with metallic/ceramic anvils, which is used for producing nanostructured materials [25], friction reaches the maximum possible level due to large asperities. Utilizing dRDAC with rough-DA will enable in situ study and optimization of occurring processes. Also, an increase in friction with rough-DA not only intensifies plastic flow but also increases pressure gradient and allows to increase in the maximum possible pressure in DAC and RDAC.

[35] Raterron, P., Detrez, F., Castelnau, O., Bollinger, C., Cordier, P., & Merkel, S. Multiscale modeling of upper mantle plasticity: from single-crystal rheology to multiphase aggregate deformation. *Physics of the Earth and Planetary Interiors*, 228, 232-243 (2014).
[36] Anderson, P.M., Hirth, J.P. & Lothe, J. Theory of dislocations. Cambridge University Press, (2017).
[37] Chen, H., Levitas, V.I. & Xiong, L. Amorphization induced by 60o shuffle dislocation pileup against different grain boundaries in silicon bicrystal under shear. *Acta Materialia*, 179, 287-295 (2019).
[38] Peng, Y., Ji, R., Phan, T., Capolungo, L., Levitas, V.I. & Xiong, L. Effect of a micro-scale dislocation pileup on the atomic-scale multi-variant phase transformation and twinning. arXiv preprint arXiv:2208.03592 (2022).
[39] Peng, Y., Ji, R., Phan, T., Gao, W., Levitas, V.I. & Xiong, L. An Atomistic-to-Microscale Computational Analysis of the dislocation pileup-induced local stresses near an interface in plastically deformed two-phase materials. *Acta Materialia*, 226, 117663 (2022).
[40] Voyiadjis, G.Z. & Yaghoobi, M. Size effects in plasticity: from macro to nano. Academic Press, (2019).
[41] Ottonello, G., Civalleri, B., Ganguly, J., Vetuschi Zuccolini, M., & Noel, Y. Thermophysical properties of the α–β–γ polymorphs of Mg2SiO4: a computational study. *Physics and Chemistry of Minerals*, 36, 87-106 (2009).
[42] Andrault, D., Bouhifd, M.A., Itie, J.P., & Richet, P. Compression and amorphization of (Mg, Fe) 2SiO4 olivines: An X-ray diffraction study up to 70 GPa. *Physics and Chemistry of Minerals*, 22, 99-107 (1995).
[43] Wu, T.C., Bassett, W.A., Burnley, P.C., Weathers M.S. Shear-promoted phase transitions in Fe2SiO4 and Mg2SiO4 and the mechanism of deep earthquakes. *Journal of Geophysical Research: Solid Earth*, 98 (B11), 19767-19776 (1993).
[44] Kawakatsu, H. & Yoshioka, S. Metastable olivine wedge and deep dry cold slab beneath southwest Japan. *Earth and Planetary Science Letters*, 303, 1-10 (2011).
[45] Billen, M.I. Deep slab seismicity limited by rate of deformation in the transition zone. *Science Advances*, 6, eaaz7692 (2020).

# Supplementary Materials for

## Plastic strain-induced olivine-ringwoodite phase transformation at room temperature: main rules and the mechanism of the deep-focus earthquake

Feng Lin[1]*, Valery I. Levitas[1, 2, 3]*, Sorb Yesudhas[1], Jesse Smith[4]

[1]Department of Aerospace Engineering, Iowa State University, Ames, Iowa 50011, USA

[2]Department of Mechanical Engineering, Iowa State University, Ames, Iowa 50011, USA

[3]Ames National Laboratory, US Department of Energy, Iowa State University, Ames, Iowa 50011, USA

[4] HPCAT, X-ray Science Division, Argonne National Laboratory, Argonne, Illinois 60439, USA

*Corresponding authors. Email: flin1@iastate.edu and vlevitas@iastate.edu

**This PDF file includes:**

> Materials and Methods
> Fig. S1
> Tables S1 to S2
> References

**Materials and Methods**

Starting materials and experiment details

The material used in this study is natural San Carlos olivine $(Mg_{0.91}Fe_{0.09})_2SiO_4$ with detailed composition measured through SEM-EDS shown in Table S1. Ground olivine powder sample is loaded in an iron gasket with an indented thickness of 150 µm and a sample chamber diameter of 200 µm. The sample is compressed and sheared using the dRDAC implemented with rough diamond anvils, following the deformation path defined in Table S2. Compression is imposed with a gas-membrane system, and shear is imposed with a servomotor within dRDAC at a controllable rotation rate. Before reaching 180° shear, where no ringwoodite phase is observed, a rotation rate of 1 rotation/min is imposed. After 180° shear, the sample is sheared at a constant rate of 0.1rotation/min. At each shear step, diffraction images were taken along the sample radius with a wavelength of 0.3445Å and 10 µm step size. Sample thickness is measured through x-ray absorption of the indented gasket area using a linear attenuation equation (Table S2). All the in-situ axial XRD experiments and x-ray absorption measurements were performed at 16-ID-B beamline at HPCAT (Sector 16) at Advanced Photon Source and recorded with a Pitalus detector. The diffraction images were converted to unrolled patterns using FIT2D software [1] and then analyzed through Rietveld refinement using MAUD software [2] to obtain the lattice parameters, volume fractions, microstrains, and crystallite sizes.

Dislocation density estimation

The crystallite sizes and microstrains extracted from the refinement using MAUD were used to estimate the dislocation density, which helps in situ tracking of the microstructure change during deformation. Dislocation density can be expressed as [3]:
$$\rho = \sqrt{\rho_c \rho_{ms}} \,. \tag{S1}$$
Here $\rho_c$ and $\rho_{ms}$ contribute to overall dislocation density from crystallite size and microstrain, respectively. Contribution from crystallite size is:
$$\rho_c = \frac{3}{d^2}. \tag{S2}$$
Where $d$ is crystallite size. Contribution from the microstrain is determined by the equation:
$$\rho_{ms} = k\varepsilon^2/b^2. \tag{S3}$$
Where $\varepsilon$ is the microstrain; $b$ is the magnitude of the Burgers vector; $k = 6\pi A(\frac{E}{G \ln(r/r_0)})$ is a material constant; $E$ and $G$ are Young's modulus and shear modulus, respectively; $A$ is a constant that lies between 2 and $\pi/2$ based on the distribution of strain; $r$ is the radius of crystallite with dislocation; $r_0$ is a chosen integration limit for dislocation core. In this study, $A = \pi/2$ for the Gaussian distribution of strain. Moduli $E$, $G$ and their pressure dependence for San Carlos olivine are taken from [4], respectively. A reasonable value of $\ln(r/r_0)$ being 4 is used [3]. Olivine has the easiest and dominant slip system of (010)[001]; thus, the length of Burger vector [001] is used [5].

**Supplementary figures**

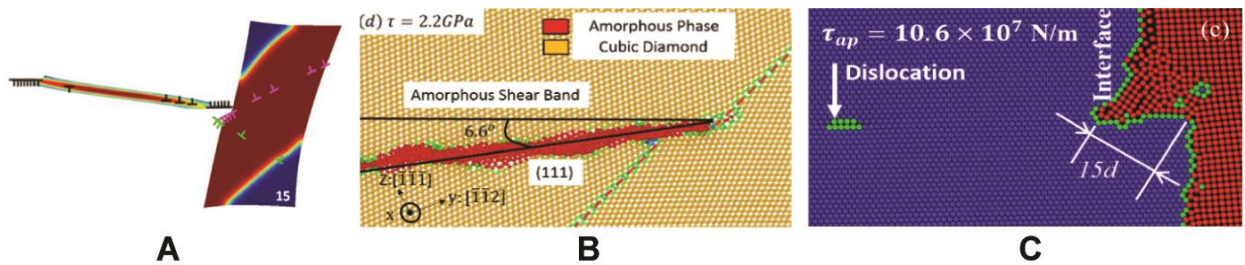

**Fig. S1. Dislocation pileups produce a step at the grain boundary or phase interface that causes a phase transformation.** (**A**) Dislocation pileup in the left grain produces step at the grain boundary and cubic to tetragonal PT and dislocation slip in the right grain. Phase-field approach results from [6]. (**B**) Dislocation pileup in the right grain produces a step at the grain boundary in Si I and amorphization in the left grain. Molecular dynamics results from [7]. (**C**) Step at the phase interface boundary consisting of 15 dislocations and causing cubic to hexagonal PT. The atomistic portion of the concurrent continuum-atomistic approach from [8]. Adopted with changes from [6-8] with permissions.

**Supplementary tables**

**Table S1.** Composition of the San Carlos olivine in this study

|            | O     | Mg    | Si    | Ca   | Cr   | Mn   | Fe   | Ni   |
|------------|-------|-------|-------|------|------|------|------|------|
| **Atom (%)**   | 56.31 | 26.53 | 14.41 | 0.03 | 0.01 | 0.04 | 2.57 | 0.11 |
| **Weight (%)** | 42.81 | 30.65 | 19.24 | 0.05 | 0.03 | 0.10 | 6.82 | 0.31 |

**Table S2.** Deformation step in the experiment and the corresponding sample thickness

| Step | Initial | Compression before 60º shear | After 60º shear | Compression before 180º shear | After 180º shear |
|------|---------|------------------------------|-----------------|-------------------------------|------------------|
| **Thickness (µm)** | 149 | 104 | 75 | 73 | 53 |
| **Step** | Compression before 270º shear | After 270º shear | After 300º shear | After 330º shear | |
| **Thickness (µm)** | 48 | 44 | 40 | 34 | |